\documentclass[conference]{IEEEtran}
\IEEEoverridecommandlockouts
\usepackage{cite}
\usepackage{amsmath,amssymb,amsfonts}
\usepackage{algorithmic}
\usepackage{graphicx}
\usepackage{textcomp}
\usepackage{xcolor}
\def\BibTeX{{\rm B\kern-.05em{\sc i\kern-.025em b}\kern-.08em
    T\kern-.1667em\lower.7ex\hbox{E}\kern-.125emX}}
\begin{document}

\title{Sum Capacity Loss Quantification With Optimal and Sub-Optimal Precoding in Heterogeneous Multiuser Channels\vspace{-7pt}}

\author{Harsh Tataria*, Mansoor Shafi$^\dagger$, and Dino Pjanić*\\
*Ericsson AB, Lund, Sweden\\
$^{\dagger}$Spark New Zealand, Wellington, New Zealand\\
e-mail: \{harsh.tataria, dino.pjanic\}@ericsson.com, and mansoor.shafi@spark.co.nz\vspace{-12pt}}
\maketitle

\begin{abstract}
We analytically approximate the expected sum capacity loss between the optimal downlink precoding technique of dirty paper coding (DPC), and the sub-optimal technique of zero-forcing precoding, for multiuser channels. We also consider the most general case of multi-stream transmission to multiple users, where we evaluate the expected sum capacity loss between DPC and block diagonalization precoding. Unlike previously, assuming heterogeneous Ricean fading, we utilize the well known affine approximation to predict the expected sum capacity difference between both precoder types (optimal and sub-optimal) over a wide range of system and propagation parameters. Furthermore, for single-stream transmission, we consider the problem of weighted sum capacity maximization, where a similar quantification of the sum capacity difference between the two precoder types is presented. In doing so, we disclose that power allocation to different users proportional to their individual weights asymptotically maximizes the weighted sum capacity. Numerical simulations are presented to demonstrate the tightness of the developed expressions relative to their simulated counterparts.  
\end{abstract}

\vspace{-8pt}
\section{Introduction}
\label{Introduction}
\vspace{-3pt}
Since the inception of massive multiple-input multiple-output (MIMO) systems, the downlink multiuser broadcast channel has, yet again, been under intense research limelight \cite{RUSEK1}. It is now well understood that dirty paper coding (DPC) -- an optimal precoding technique pioneered by Costa in \cite{COSTA1}, achieves the \emph{capacity region} of the multiuser broadcast channel \cite{RUSEK1}. Nonetheless, practical implementation of DPC requires substantial complexity for ordered decoding and interference cancellation with precise channel state information at both the cellular base station (BS) and user terminals \cite{RUSEK1}. This has paved the way for reduced complexity, yet sub-optimal, linear precoding techniques such as zero-forcing (ZF) and block diagonalization (BD) for multiuser transmission with single and/or multiple streams. Due to their sub-optimal nature, linear precoding techniques incur an absolute expected (average) sum capacity penalty (a.k.a. expected sum capacity offset) relative to DPC, reducing the achievable performance \cite{JINDAL1}. With massive MIMO systems, linear precoding can achieve up to 98\% of DPC performance if the number of BS antennas exceeds at least 10$\times$ the total number of user antennas \cite{RUSEK1}. 

Despite the many research advances on multiuser broadcast channels (see e.g., \cite{NAM2,CAIRE1,LOZANO1} and references therein), understanding and quantifying the fundamental \emph{difference} in the expected sum capacity with optimal and sub-optimal precoding techniques remains a sparsely investigated area. \emph{The focus of the paper is to close this research gap.} The authors of \cite{LOZANO1,LEE1} present some preliminary results on this topic for point-to-point and multiuser MIMO systems for simple uncorrelated Rayleigh fading channels. Nevertheless, such analysis do not capture, to the full extent, the heterogeneity present in multiuser channels due to wide ranging propagation conditions. In reality, some users may experience the presence of dominant line-of-sight (LOS) components, while others may be experience heavy non-LOS (NLOS) conditions. In contrast to \cite{LOZANO1,LEE1}, we consider the general case of a heterogeneous Ricean fading channel, where each terminal has a unique Ricean $K$-factor and LOS component steering angle. For single-stream transmission, we analyze the expected sum capacity loss between DPC and ZF, and for multi-stream transmission, we consider BD precoding expected sum capacity performance relative to DPC over a wide range of operating signal-to-noise ratios (SNRs), BS/user antenna numbers, and Ricean $K$-factors. Our analysis leverages the affine approximation to sum capacity developed in \cite{SHAMAI1}, which states that $C(\textrm{SNR})\approx\mathcal{S}_{\infty}[\hspace{1pt}\log_{2}(\textrm{SNR})-\mathcal{L}_{\infty}]+\mathcal{O}(1)$. Here, $\mathcal{S}_{\infty}[\cdot]$ denotes the \emph{multiplexing gain} (additional bits/sec/Hz for every 3 dB increase in SNR), $\mathcal{L}_{\infty}$ denotes the \emph{sum capacity offset} and $\mathcal{O}(1)$ represents an order one term. Though this approximation is \emph{exact} at asymptotically high SNRs (see \cite{LOZANO1,LEE1}), it is seen to provide tight results over a wide range of SNRs. When applying this approximation to the DPC and ZF sum capacity, the same $\mathcal{S}_{\infty}$ terms are obtained, with different $\mathcal{L}_{\infty}$ terms allowing us to characterize the difference in sum  capacity between optimal and sub-optimal precoding. By averaging the per-channel realization sum capacity offset over the myriad of Ricean fading, we are able to derive simple, yet accurate, expressions for the expected sum capacity difference as a function of system and propagation parameters. While the studies in \cite{JINDAL2,SHEN1} have analyzed the \emph{ratio} between sum capacity obtained with non-linear and linear precoding for Rayleigh fading multiuser channels, we study the \emph{absolute difference} between these to quantify the sum capacity loss. 

In addition the expected sum capacity loss quantification, we also investigate the weighted sum capacity maximization problem using DPC and ZF for single-stream transmission providing simple expressions for the capacity offsets. We show that the weighted sum capacity is maximized at asymptotically high SNRs by allocating power in direct proportion to the user weights. This result generalizes the well understood property that equal power allocation (e.g., across users and fading states) asymptotically maximizes the sum capacity of heterogeneous multiuser channels. 

\textbf{Notation.} Upper and lower boldface letters represent matrices and vectors. The $M\times{}M$ identity matrix is denoted as $\mathbf{I}_{M}$, while the $M\times{}M$ zero matrix is denoted by $\mathbf{0}_{M}$. The $(i,j)$-th entry of the matrix $\textbf{X}$ is denoted by $(\mathbf{X})_{i,j}$, while the transpose, Hermitian transpose, inverse and trace operators are denoted by $(\cdot)^{T}$, $(\cdot)^{H}$, $(\cdot)^{-1}$ and $\textrm{Tr}[\cdot]$, respectively. Moreover, $||\cdot||$ and  $|\cdot|$ denote the Euclidian vector norm and determinant operator of a matrix. We use $\mathbf{h}\sim\mathcal{CN}(\mathbf{m},\mathbf{R})$ to denote a complex Gaussian distribution for $\mathbf{h}$ with mean $\mathbf{m}$ and covariance matrix $\mathbf{R}$. Finally, we write $\mathbb{E}\{\cdot\}$ to denote the statistical expectation, and $\textrm{max}\{f(x)\}$ to denote the maximum value of the function $f(x)$, where $x$ is a scalar value. 

\vspace{-2pt}
\section{System Model}
\label{SystemModel}
\vspace{-1pt}
We consider a downlink of an $L$ user MIMO broadcast channel, in which the BS has a uniform linear array (ULA) of $M$ electronically steerable antenna elements. For maximizing generality, each user terminal is equipped with $N$ antennas, where $M\geq{}LN$. The $N\times{}1$ received signal observed at the $N$ antennas of the $\ell$-th user terminal can then be written as
\vspace{-2pt}
\begin{equation}
    \label{receivedsignal}
    \mathbf{y}_{\ell}=\rho^{\frac{1}{2}}\hspace{1pt}\mathbf{H}_{\ell}\hspace{1pt}\mathbf{x}+\mathbf{n}_{\ell},
    \vspace{-5pt}
\end{equation}
where $\rho$ denotes the average transmit power from the BS and $\mathbf{H}_{\ell}$ denotes the $N\times{}M$ downlink propagation channel matrix to terminal $\ell$ from the BS encapsulating the effects of small-scale fading (discussed later in the text). Additionally, $\mathbf{x}$ is the $M\times{}1$ transmitted signal vector from the $M$ BS antennas having the total average power constraint $\textrm{Tr}[\mathbb{E}\{\mathbf{x}\mathbf{x}^H\}]\leq{}\rho$, and $\mathbf{n}_{\ell}$ is the $N\times{}1$ additive white Gaussian noise with unit variance per-component, such that $\mathbb{E}\{\mathbf{n}_{\ell}\mathbf{n}_{\ell}^{H}\}=\mathbf{I}_{N}$. Without loss of generality, we assume the same noise covariance matrix for each user. To this end, we define the operating SNR as the ratio of the average transmit power to the per-user noise variance, i.e., $\textrm{SNR}=\rho$. To facilitate the analysis of the expected sum capacity loss between DPC and ZF precoding, as well as BD precoding, we assume that the BS has knowledge of all propagation channel matrices, and each user terminal has knowledge of its individual propagation channel matrix. At first glance, this assumption may come across as rather idealistic. However, such an assumption is necessary to make analytical progress, and the results derived here can be regarded as a useful upper bound in performance relative to what is obtainable with estimated channels. We denote the \emph{composite} $LN\times{}M$ channel by  $\mathbf{H}^{H}=[\mathbf{H}_{1}^{H},\mathbf{H}_{2}^{H},\dots,\mathbf{H}_{L}^{H}]$, which can be further decomposed as  $\mathbf{H}^{H}=[\hspace{1pt}(\mathbf{h}_{1,1}^{H}\hspace{3pt}\mathbf{h}_{1,2}^{H},\dots,\mathbf{h}_{1,N}^{H})\dots{}(\mathbf{h}_{L,1}^{H},\mathbf{h}_{L,2}^{H},\dots{},\mathbf{h}_{L,N}^{H})\hspace{1pt}]$, where $\mathbf{h}_{\ell,n}$ is the $1\times{}M$ $n$-th row vector of $\mathbf{H_{\ell}}$. The small-scale fading matrix for user $\ell$, $\mathbf{H}_{\ell}$, follows a Ricean distribution, which can be expressed as \cite{TATARIA2} 
\vspace{-2pt}
\begin{equation}
    \label{channeltoUEl}
    \mathbf{H}_{\ell}=\bar{\mathbf{H}}_{\ell}\hspace{2pt}\sqrt{\frac{\kappa_{\ell}}{
    \kappa_{\ell}+1}}\hspace{2pt}+\hspace{2pt}\tilde{\mathbf{H}}_{\ell}\hspace{2pt}\sqrt{\frac{1}{1+\kappa_{\ell}}}. 
    \vspace{-2pt}
\end{equation}
Here, $\bar{\mathbf{H}}_{\ell}$ is the deterministic component modelling the presence of LOS or dominant components in the propagation channel, with its  $(n,m)$-th entry governed by 
\vspace{-5pt}
\begin{equation}
    \label{LOSpropagationmatrixtoUEl}
    \left(\hspace{1pt}\bar{\mathbf{H}}\hspace{1pt}\right)_{n,m}=\textrm{exp}\left(-j\left(n-1\right)\frac{2\pi{}d}{\tilde{\lambda}}\sin\left(\phi_{m}\right)\right). 
    \vspace{-4pt}
\end{equation}
We note that $d$ represents the inter-element spacing between successive antenna elements of the ULA, $\tilde{\lambda}$ denotes the carrier wavelength and $\phi_{m}$ is the angle-of-departure (AOD) from the $m$-th antenna element at the BS. Moreover, $\kappa_{\ell}$ denotes the Ricean $K$-factor of terminal $\ell$ governing the ratio of LOS to NLOS power \cite{TATARIA2}. The NLOS, i.e., purely stochastic component of the channel for user $\ell$ is modelled via the $N\times{}M$ matrix, $\tilde{\mathbf{H}}_{\ell}$, the entries of which are independent and identically distributed Gaussian random variables with zero-mean and unit variance. Thus, $\tilde{\mathbf{H}}_{\ell}\sim\mathcal{CN}(0,\mathbf{I}_{M})$. To maximize clarity, exact parameterization of above quantities is given in Sec.~\ref{NumericalResults}, where further  details are presented. In what follows, we present the methodology for the expected sum capacity loss quantification between DPC and ZF, as well as BD precoding.

\vspace{-3pt}
\section{Expected Sum Capacity Loss Quantification Methodology}
\label{CapacityLossQuantificationMethodology}
\vspace{-1pt}
\subsection{Dirty Paper Coding Sum Capacity Analysis}
\label{DirtyPaperCodingCapacityAnalysis}
\vspace{-1pt}
The DPC sum capacity hits the multiuser broadcast channel capacity region, which can be expressed from the landmark results of \cite{VISHWANATH1} on the broadcast to multiple access channel \emph{duality}. This allows us to express the DPC sum capacity as 
\vspace{-4pt}
\begin{equation}
    \label{DPCCapacity}
    C_{\textrm{DPC}}\hspace{-1pt}\left(\mathbf{H},\hspace{-1pt}\rho\right)=\hspace{-7pt}\underset{\sum\nolimits_{\ell=1}^{L}\hspace{-2pt}\textrm{Tr}\left[\mathbf{Q}_{\ell}\right]\leq{}\rho}{\textrm{max}}\hspace{-2pt}\left\{\hspace{-1pt}\log_{2}\left|\hspace{1pt}\mathbf{I}_{M}\hspace{-2pt}+\hspace{-2pt}\sum\limits_{\ell=1}^{L}\mathbf{H}_{\ell}^{H}\hspace{-1pt}\mathbf{Q}_{\ell}\mathbf{H}_{\ell}\right|\hspace{1pt}\right\}\hspace{-1pt}, 
    \vspace{-2pt}
\end{equation}
where $\mathbf{Q}_{\ell}$ is the $N\times{}N$ covariance matrix of the \emph{dual} multiple access channel. It is well known that no closed-form solution to \eqref{DPCCapacity} exists, yet it has been shown in \cite{VISHWANATH1} that $C_{\textrm{DPC}}(\mathbf{H},\rho)$ converges in the absolute sense to the capacity of a point-to-point MIMO propagation channel with propagation matrix  $\mathbf{H}$, when $M\geq{}LN$. As $\rho\rightarrow\infty$ \cite{CAIRE1}, 
\vspace{-2pt}
\begin{equation}
\label{DPCCapacityApproximation1}
    \underset{\rho\rightarrow{}\infty}{\textrm{lim}}\left\{C_{\textrm{DPC}}\left(\mathbf{H},\rho\right)-\log_{2}\left|\hspace{1pt}\mathbf{I}_{M}+\frac{\rho}{LN}\mathbf{H}^{H}\mathbf{H}\hspace{1pt}\right|\hspace{1pt}\right\}=0. 
    \vspace{-2pt}
\end{equation}
One can discover a corollary of this result if the covariance matrices are designed such that $\mathbf{Q}_{\ell}=(\rho/LN)\mathbf{I}_{N}$, leading to an asymptotically optimal solution to \eqref{DPCCapacity}. As such, an affine approximation (stated in the Sec.~\ref{Introduction}) can be applied, such that the DPC sum capacity can be approximated as 
\vspace{-2pt}
\begin{equation}
    \label{DPCCapacityApproximation2}
    C_{\textrm{DPC}}\left(\mathbf{H},\rho\right)\approx{}L\hspace{-1pt}N\hspace{-1pt}
    \left[\hspace{1pt}\log_{2}\rho\hspace{-1pt}-\hspace{-1pt}\log_{2}\hspace{-1pt}L\hspace{-1pt}N\hspace{1pt}\right]\hspace{-1pt}+\hspace{-1pt}
    \log_{2}\left|\mathbf{H}^{H}\mathbf{H}\right|. 
    \vspace{-3pt}
\end{equation}
While this result becomes exact in the limit of high SNRs, we note that the high SNR sum capacity primarily depends on the product of the number of user terminals ($L$), and the number of receive antennas per-user ($N$). The effects of other parameters such as the number of BS antennas ($M$), the Ricean $K$-factors and LOS steering angles are captured in the $|\mathbf{H}^{H}\mathbf{H}|$ term. 
\vspace{-4pt}
\subsection{Linear/Sub-Optimal Precoding Sum Capacity Analysis}
\label{LinearPrecodingCapacityAnalysis}
\vspace{-2pt}
We now evaluate the affine approximation to the linear precoding sum capacity using ZF and BD processing, and quantify the sum capacity offset relative to DPC. With linear precoding, the transmitted signal vector, $\mathbf{x}$, can be written as 
\vspace{-4pt}
\begin{equation}
    \label{linearprecodingcapacity1}
    \mathbf{x}=\sum\limits_{\ell=1}^{L}\mathbf{W}_{\hspace{-1pt}\ell}
    \hspace{2pt}\mathbf{s}_{\ell}. 
    \vspace{-3pt}
\end{equation}
Here, $\mathbf{s}_{\ell}$ is the $N\times{}1$ vector of data symbols intended for user terminal $\ell$, and $\mathbf{W}_{\ell}$ is the $M\times{}N$ precoding matrix for user $\ell$. The received signal at user $\ell$ can then be written as 
\vspace{-7pt}
\begin{equation}
    \label{receivedsignalUElwithlinearprecoding1}
    \mathbf{y}_{\ell}=\rho^{\frac{1}{2}}\hspace{1pt}\mathbf{H}_{\ell}\mathbf{W}_{\hspace{-1pt}\ell}\hspace{2pt}
    \mathbf{s}_{\ell}+\rho^{\frac{1}{2}}\hspace{2pt}\sum\limits_{\substack{i=1\\i\neq{}\ell}}^{L}
    \hspace{1pt}\mathbf{H}_{\ell}\mathbf{W}_{\hspace{-1pt}i}\hspace{1pt}\mathbf{s}_{i}+\mathbf{n}_{\ell},
    \vspace{-5pt}
\end{equation}
where the first term denotes the desired signal power, while the second and third terms denote the multiuser interference (including multi-stream interference) and additive white Gaussian noise. In the case of BD precoding (multi-stream transmission), multiuser interference is \emph{nulled} by designing the individual user precoding matrices such that $\mathbf{H}_{\ell}\mathbf{W}_{i}=\mathbf{0}_{N}$, $\forall{}i\neq{}\ell\in{}[1,L]$. On the otherhand, for single-stream transmission, ZF precoders can be designed such that $\mathbf{h}_{\ell,n}\mathbf{w}_{i,p}=0$, $\forall{}i\neq{}\ell\in[1,L]$ and $\forall{}n,p\in[1,N]$. Here, $\mathbf{w}_{i,p}$ denotes the $p$-th column vector of $\mathbf{W}_{\hspace{-1pt}i}$. Keeping this in mind, the received signal at the $n$-th antenna of the $\ell$-th user terminal is given by 
\vspace{-2pt}
\begin{equation}
    \label{receivedsignalUElwithlinearprecoding2}
    y_{\ell,n}=\mathbf{h}_{\ell,n}\mathbf{w}_{\ell,n}\mathbf{s}_{\ell}+n_{\ell,n},\hspace{8pt}n=1,2,\dots{},N. 
    \vspace{-3pt}
\end{equation}
As such, ZF precoding converts the system into $LN$ parallel channels with an \emph{equivalent} channel entry $g_{\ell,n}=\mathbf{h}_{\ell,n}\mathbf{w}_{\ell,n}$.\footnote{We exercise a slight abuse of notation here and denote $n_{\ell,n}$ as the additive white Gaussian noise to the observation received at the $n$-th \emph{receive antenna} of the $\ell$-th \emph{user terminal}. We caution the reader to be mindful of this.} To \emph{maximize} sum capacity, optimal power allocation across the $LN$ channels should take place. Doing this would yield
\vspace{-2pt}
\begin{equation}
    \label{ZFCapacity1}
    C_{\textrm{ZF}}\hspace{-1pt}\left(\mathbf{H},\rho\right)\hspace{1pt}=\hspace{-6pt}\underset{\sum\nolimits_{\ell=1}^{L}\sum\nolimits_{n=1}^{N}\rho_{\ell,n}\leq{}\rho}{\textrm{max}}\hspace{3pt}\sum\limits_{\ell=1}^{L}\sum\limits_{n=1}^{N}\log_{2}1+\delta_{\ell,n},  
    \vspace{-3pt}
\end{equation}
where $\delta_{\ell,n}=\rho_{\ell,n}\left|g_{\ell,n}\right|^{2}$. Since uniform power allocation coefficients converge to optimal power allocation at asymptotically high SNRs \cite{JINDAL1}, applying the affine approximation, 
\vspace{-5pt}
\begin{equation}
    \label{ZFCapacity2}
     C_{\textrm{ZF}}\hspace{-1pt}\left(\mathbf{H},\rho\right)\approx{}
     \alpha+
     \log_{2}\hspace{2pt}\prod\limits_{\ell=1}^{L}\prod\limits_{n=1}^{N}\hspace{1pt}
    \left|\hspace{1pt}g_{\ell,n}\right|^{2}, 
    \vspace{-2pt}
\end{equation}
where $\alpha=L\hspace{-1pt}N 
    [\hspace{1pt}\log_{2}\rho\hspace{-1pt}-\hspace{-1pt}\log_{2}\hspace{-1pt}L\hspace{-1pt}N]$. The expression in \eqref{ZFCapacity2} has the same form of that derived for DPC sum capacity in \eqref{DPCCapacityApproximation2}, with the exception of the second term on the right-hand side (RHS) of \eqref{ZFCapacity2}, which is specific to ZF processing. As such, we define the asymptotic sum capacity \emph{loss} as the \emph{difference} between the DPC and ZF capacity. This can be expressed as 
    \begin{align}
        \nonumber
        C_{\textrm{loss}}^{\textrm{DPC-ZF}}\left(\mathbf{H}\right)=&\underset{\rho\rightarrow{}\infty}
        {\textrm{lim}}\left\{C_{\textrm{DPC}}\left(\mathbf{H},\rho\right)-C_{\textrm{ZF}}\left(\mathbf{H},\rho\right)\right\}\\
        \label{capacitylossanalysis1}
        =\hspace{2pt}&
        \log_{2}\frac{\left|\mathbf{H}^{H}\mathbf{H}\right|}{\prod\limits_{\ell=1}^{L}\hspace{-1pt}\prod\limits_{n=1}^{N}\left|g_{\ell,n}\right|^{2}}. 
    \end{align}
While the above metric denotes the sum capacity loss per-realization of $\mathbf{H}$, we can also determine the \emph{expected} capacity loss across the myriad of small-scale fading. That is, 
\vspace{-3pt}
\begin{equation}
    \label{capacitylossanalysis2}
    \mathbb{E}\{C_{\textrm{loss}}^{\textrm{DPC-ZF}}\left(\mathbf{H}\right)\}\approx\mathbb{E}\left\{
    \log_{2}\frac{1}{\gamma}\left|\mathbf{H}^{H}\mathbf{H}\hspace{1pt}\right|\right\}, 
    \vspace{-2pt}
\end{equation}
where $\gamma=\prod\nolimits_{\ell=1}^{L}\prod\nolimits_{n=1}^{N}|g_{\ell,n}|^{2}$. In general, it is extremely difficult (if not intractable) to exactly analyze the expected value of \eqref{capacitylossanalysis1}, due to the heterogeneity present in the composite multiuser channel. To this end, we employ the well characterized (see e.g., \cite{TATARIA3,ZHANG1} and references therein) first-order Laplace approximation, which takes the form as shown in \eqref{capacitylossanalysis2}. The accuracy of such an approximation relies on the denominator of \eqref{capacitylossanalysis2} having a small \emph{second moment} relative to its \emph{first moment}. This can be seen via application of a multivariate Taylor series expansion of \eqref{capacitylossanalysis2} around its first moment value. The approximation in \eqref{capacitylossanalysis2} has shown to give tight results when $M$ and $L$ start to grow (the case for massive MIMO), since the implicit averaging in the denominator of \eqref{capacitylossanalysis2} gives rise to the required variance reduction \cite{TATARIA3,ZHANG1}. The result in \eqref{capacitylossanalysis2} can equivalently be written as 
\vspace{-1pt}
\begin{equation}
    \label{capacitylossanalysis3}
    \mathbb{E}\{C_{\textrm{loss}}^{\textrm{DPC-ZF}}\left(\mathbf{H}\right)\}\approx\mathbb{E}\left\{
    \log_{2}\frac{1}{\gamma}\right\}+\mathbb{E}\left\{\log_{2}\left|\mathbf{H}^{H}\mathbf{H}\hspace{1pt}\right|\right\}. 
    \vspace{-1pt}
\end{equation}
Analyzing the second term on the RHS of \eqref{capacitylossanalysis3}, we can observe that $\mathbf{H}^H\mathbf{H}$ follows an \emph{uncorrelated non-central Wishart distribution} denoted by $\mathbf{H}^H\mathbf{H}\sim\mathcal{W}_{L}(M,\mathbf{P},\mathbf{\Sigma})$ \cite{TULINO1}. Here by definition, we note that $\mathbf{P}=\bar{\mathbf{H}}\hspace{1pt}[\hspace{1pt}\boldsymbol{\kappa}(\boldsymbol{\kappa}+\mathbf{I}_{L})^{-1}]^{1/2}$ is the \emph{mean matrix} of $\mathbf{H}$ with $\bar{\mathbf{H}}$ capturing each user's LOS channel matrix. Additionally, $\boldsymbol{\kappa}$ is an $L\times{}L$ diagonal matrix containing $K$-factors for all users, and $\mathbf{\Sigma}=(\boldsymbol{\kappa}+\mathbf{I}_{L})^{-1}$ is the \emph{covariance matrix of the row vectors} of $\mathbf{H}$. Leveraging the results in \cite{KIM1} on the expected log-determinants of non-central Wishart matrices, the second term of the RHS of \eqref{capacitylossanalysis3} is given by 
\vspace{-2pt}
\begin{equation}
\label{capacitylossanalysis4}
    \mathbb{E}\left\{\log_{2}\left|\mathbf{H}^{H}\mathbf{H}\hspace{1pt}
    \right|\right\}=
    \sum\limits_{\ell=1}^{L\hspace{-1pt}N}\Delta_{M}\left(\lambda_{\ell}\right), 
    \vspace{-2pt}
\end{equation}
where $\lambda_{\ell}$ is the $\ell$-th eigenvalue of $\bar{\mathbf{H}}^{H}\bar{\mathbf{H}}$. Note that $\Delta_{M}(\lambda_{\ell})$ is defined in \eqref{capacitylossanalysis5} on the top of the following page for reasons of space, where $\textrm{Ei}(\cdot)$ denotes the exponential integral function and $(\cdot)\hspace{1pt}!$ denotes the factorial operation \cite{ISR1}. 
\begin{figure*}[!t]
\begin{equation}
\label{capacitylossanalysis5}
    \Delta_{M}\left(\lambda_{\ell}\right)=\log_2{\lambda_{\ell}}-\textrm{Ei}\left(
    -\lambda_{\ell}\right)+\sum\limits_{\ell=1}^{M-1}\left(\frac{-1}
    {\lambda_{\ell}}\right)^{\ell}
    \left[\hspace{1pt}\textrm{exp}\left(\lambda_{\ell}\right)\left(\ell-1\right)\hspace{1pt}!-
    \frac{\left(M-1\right)\hspace{1pt}!}{\ell\left(M-1-\ell\right)\hspace{1pt}!}\right]. 
\end{equation}
\hrulefill
\vspace{-10pt}
\end{figure*}
 In contrast to this, the first term of the RHS of \eqref{capacitylossanalysis3} is obtained by approximating the structure of $\mathbf{H}^{H}\mathbf{H}$ by the often used (see e.g., \cite{TATARIA2,ZHANG1}) \emph{uncorrelated central Wishart distribution} with a shift in the covariance matrix $\mathbf{\Sigma}$ (following the methodology in \cite{ZHANG1})\footnote{Detailed investigation into the accuracy of the quoted approximation from non-central to central Wishart matrices can be found in \cite{ZHANG1}.} to 
\begin{align}
    \nonumber
    \widehat{\mathbf{\Sigma}}=&\left(\boldsymbol{\kappa}+\mathbf{I}_{L}\right)^{-1}+\frac{1}{M}\Bigg[
    \sqrt{\boldsymbol{\kappa}\left(\boldsymbol{\kappa}+\mathbf{I}_{L}\right)^{-1}}\hspace{2pt}\\[-2pt]
    \label{capacitylossanalysis6}
    &\hspace{20pt}\times{}\bar{\mathbf{H}}^{H}\bar{\mathbf{H}}\hspace{3pt}\sqrt{\boldsymbol{\kappa}\left(\boldsymbol{\kappa}+\mathbf{I}_{L}\right)^{-1}}\hspace{1pt}\Bigg]. \\[-23pt]
    \nonumber&
\end{align}
Considering the above, one can recognize that $|g_{\ell,n}|^{2}$ is a \emph{chi-squared random variable} with $M-LN+1$ degrees-of-freedom. Applying the standard probability density of the chi-squared random variable together with some straightforward algebraic manipulations, we can express the first term of the RHS of \eqref{capacitylossanalysis3} as follows; where 
\vspace{-3pt}
\begin{equation}
    \label{capacitylossanalysis7}
    \mathbb{E}\left\{\log_{2}\frac{1}{\gamma}\right\}\hspace{1pt}\approx{}\hspace{1pt}LN\left[\log_{2}\hspace{-1pt}
    \frac{\left(\widehat{\mathbf{\Sigma}}^{-1}\right)_{\left(\ell,n\right),\left(\ell,n\right)}}{M-LN}\right]\hspace{2pt}. 
    \vspace{-5pt}
\end{equation}
\emph{Adding the result in \eqref{capacitylossanalysis7} with \eqref{capacitylossanalysis4} yields the expected sum capacity loss between DPC and ZF precoding as stated in \eqref{capacitylossanalysis3}.} We note that this is a rather simple result for an extremely general and complex scenario of multiuser channels having user specific heterogeneity and parameters. Due to aforementioned mathematical difficulties, analysis of this type has been missing from the literature. In Sec.~\ref{NumericalResults}, we evaluate the accuracy of the derived expected sum capacity loss by comparing it to its simulated counterpart for a wide range of system and propagation parameters. 

Similar to ZF precoding, with BD precoding, $L$ parallel channels are formed with $N\times{}N$ \emph{equivalent channel} matrices $\mathbf{G}_{\ell}=\mathbf{H}_{\ell}\mathbf{W}_{\ell}$, $\forall{}\ell=1,2,\dots,L$. To this end, the BD sum capacity over all $L$ users can be written as \cite{JINDAL1,JINDAL2}
\vspace{-5pt}
\begin{equation}
    \label{BDCapacityAnalysis1}
    C_{\textrm{BD}}\left(\mathbf{H},\rho\right)=\hspace{-4pt}\underset{\sum\nolimits_{\ell=1}^{L}
    \textrm{Tr}\left(\mathbf{Q}_{\ell}\right)\leq{}\rho}{\textrm{max}}\hspace{1pt}
    \sum\limits_{\ell=1}^{L}\hspace{1pt}
    \log_{2}\left|\hspace{1pt}\mathbf{I}_{N}\hspace{-2pt}+\hspace{-2pt}
    \mathbf{G}_{\ell}^{H}\mathbf{Q}_{\ell}\hspace{1pt}\mathbf{G}_{\ell}\right|
    \hspace{-1pt}, 
    \vspace{-2pt}
\end{equation}
and the sum capacity can be approximated under the same constraints as for ZF processing via the affine approximation as 
\vspace{-6pt}
\begin{equation}
    \label{BDCapacityAnalysis2}
    C_{\textrm{BD}}\left(\mathbf{H},\rho\right)\approx
    \alpha+\log_{2}\hspace{1pt}\prod\limits_{\ell=1}^{L}\left|\hspace{1pt}
    \mathbf{G}_{\ell}^{H}\mathbf{G}_{\ell}\right|. 
    \vspace{-4pt}
\end{equation}
Note that $\alpha$ in \eqref{BDCapacityAnalysis2} is as defined after \eqref{ZFCapacity2}. 
Similar to \eqref{capacitylossanalysis1}, the asymptotic loss from the instantaneous DPC sum capacity relative to that achieved by BD precoding can be expressed as 
\vspace{-5pt}
\begin{equation}
    \label{BDCapacityAnalysis3}
    C_{\textrm{loss}}^{\textrm{DPC-BD}}\left(\mathbf{H}\right)=\underset{\rho\rightarrow{}\infty}{\textrm{lim}}\left\{C_{\textrm{DPC}}\left(\mathbf{H},\rho\right)-C_{\textrm{BD}}\left(\mathbf{H},\rho\right)\right\}. 
    \vspace{-2pt}
\end{equation}
Similar to the ZF case, the \emph{expected loss} across the ensemble of small-scale fading can then be evaluated by analyzing the statistical expectation term,  $\mathbb{E}\{C_{\textrm{loss}}^{\textrm{DPC-BD}}(\mathbf{H})\}$. Following a similar methodology to the ZF analysis, one can derive approximation to  $\mathbb{E}\{C_{\textrm{loss}}^{\textrm{DPC-BD}}(\mathbf{H})\}$. Due to space reasons, we omit its detailed analysis here and note that the full derivation will feature in the upcoming journal version of the paper. In Sec.~\ref{NumericalResults}, the sum capacity performance with BD precoding relative to ZF and DPC is assessed for multi-stream transmission scenarios. 

In the section which follows, for single-stream, single receive antenna per-user scenarios, we analyze the maximization of the sum capacity with power allocation to each user.

\section{Weighted Sum Capacity Maximization}
\label{WeightedCapacityMaximization}
In this section, we generalize the earlier discussed sum capacity to weighted sum capacity maximization for the case when each user has a single antenna, i.e., $N=1$. As such, the downlink propagation channel to user $\ell$ is a $1\times{}M$ row vector, denoted by $\mathbf{h}_{\ell}$. We first show that allocating power in proportion to the user weights is asymptotically optimal, and use this result to compute the associated sum capacity offsets. Without loss of generality, we assume that user weights, denoted by $\mu_{1}, \mu_{2},\dots{},\mu_{L}$, are in \emph{descending order}, i.e., $\mu_{1}\geq{}\mu_{2}\geq{},\dots,\geq{}\mu_{L}\geq{}0$ with $\sum\nolimits_{\ell=1}^{L}\mu_{\ell}=1$. The maximum weighted sum capacity problem for DPC, defined as the \emph{maximum} of $\sum\nolimits_{\ell=1}^{L}\mu_{\ell}\hspace{1pt}C_{\ell}$ over the \emph{capacity region}, which can be written as in \eqref{OptimalPowerAnalysis1} on top of the following page for reasons of space. Note that here $C_{\ell}$ denotes the $\ell$-th user's individual capacity.  
\begin{figure*}
\begin{equation}
    \label{OptimalPowerAnalysis1}
    C_{\textrm{DPC}}\left(\mu,\mathbf{H},\rho\right)=\underset{\sum\nolimits_{\ell=1}^{L}\rho_{\ell}\leq\rho}{\textrm{max}}\hspace{3pt}
    \sum\limits_{\ell=1}^{L}\mu_{\ell}\left[\hspace{2pt}\log_{2}1+\rho_{\ell}\hspace{2pt}\mathbf{h}_{\ell}\left(\mathbf{S}^{\hspace{1pt}\left(\ell-1\right)}\right)^{\hspace{-1pt}-1}\hspace{-1pt}\mathbf{h}_{\ell}^{H}\hspace{1pt}\right]. 
\end{equation}
\hrulefill
\vspace{-10pt}
\end{figure*}
In \eqref{OptimalPowerAnalysis1}, $\mathbf{S}^{(\ell-1)}=\mathbf{I}_{M}+\sum\nolimits_{j=1}^{\ell-1}\rho_{j}\hspace{1pt}\mathbf{h}_{j}^{H}\hspace{1pt}\mathbf{h}_{j}$. The result which follows, shows that if we limit ourselves to linear power allocation policies, then the maximization problem of  \eqref{OptimalPowerAnalysis1} can be decoupled at higher SNRs. Following the arguments in \cite{JINDAL1,JINDAL2}, with $M\geq{}L$, for any $\beta_{\ell}=\rho_{\ell}/\rho>0$, $\forall\ell=1,2,\dots{},L$ with condition $\sum\nolimits_{\ell=1}^{L}\beta_{\ell}=1$, we can write 
\vspace{-5pt}
\begin{align}
    \nonumber
    &\underset{\rho\rightarrow{}\infty}
    {\textrm{lim}}\left\{\sum\limits_{\ell=1}^{L}
    \mu_{\ell}\left[\log_{2}1+\rho\hspace{1pt}\beta_{\ell}\hspace{2pt}\mathbf{h}_{\ell}\hspace{1pt}\left(\mathbf{S}^{(\ell-1)}\right)^{-1}\mathbf{h}_{\ell}^{H}\right]\right.\\[-3pt]
    \label{OptimalPowerAnalysis2}
    &\left.\hspace{35pt}-\sum\limits_{\ell=1}^{L}\mu_{\ell}\left(\log_2
    1+\beta_{\ell}\rho\left\|\mathbf{f}_{\ell}\right\|^{2}\right)\right\}=0, \\[-22pt]
    &\nonumber
\end{align}
where $\mathbf{f}_{\ell}$ is the \emph{projection} of $\mathbf{h}_{\ell}$ on to the \emph{matrix null space} of $\{\mathbf{h}_{1},\mathbf{h}_{2},\dots{},\mathbf{h}_{\ell-1}\}$. That is, instead of solving \eqref{OptimalPowerAnalysis1} directly, the following optimization setup will yield an asymptotically identical solution. That is, 
\vspace{-5pt}
\begin{equation}
    \label{OptimalPowerAnalysis3}
    \underset{\sum\nolimits_{\ell=1}^{L}\rho_{\ell}\leq{}\rho}{\textrm{max}}\sum\limits_{\ell=1}^{L}\hspace{1pt}\mu_{\ell}\left(\log_{2}1+\rho_{\ell}\left\|\mathbf{f}_{\ell}
    \right\|^{2}\right). 
    \vspace{-2pt}
\end{equation}
The Karush–Kuhn–Tucker conditions (see e.g., \cite{JINDAL1,LEE1}) to the problem presented in \eqref{OptimalPowerAnalysis3} yields the solution 
\vspace{-6pt}
\begin{equation}
    \label{OptimalPowerAnalysis4}
    \rho_{\ell}*=\mu_{\ell}\hspace{1pt}\rho{}+\mu_{\ell}\left(
    \sum\limits_{\substack{i=1\\i\neq{}\ell}}^{L}\frac{1}{||\mathbf{f}_{i}||^{2}}\right)-\frac{1}{||\mathbf{f}_{\ell}||^{2}}, 
    \vspace{-5pt}
\end{equation}
for $\ell=1,2,\dots{},L$. Therefore, at high SNRs, we have 
\vspace{-3pt}
\begin{equation}
    \label{OptimalPowerAnalysis5}
    \rho_{\ell}*=\mu_{\ell}\rho+\mathcal{O}
    \left(1\right). 
    \vspace{-4pt}
\end{equation}
Since the order one term leads to a vanishes with SNR, allocating power according to $\rho_{\ell}=\rho\hspace{1pt}\mu_{\ell}$,  $\forall{}\ell=1,2,\dots{}L$ maximizes \eqref{OptimalPowerAnalysis4}. Meanwhile, the weighted sum capacity with ZF processing is given by \cite{LEE1}
\vspace{-6pt}
\begin{equation}
    \label{OptimalPowerAnalysis6}
    C_{\textrm{ZF}}\left(\mu,\mathbf{H},\rho\right)=
    \underset{\sum\nolimits_{\ell=1}^{L}\rho_{\ell}<\rho}
    {\textrm{max}}\hspace{2pt}\sum\limits_{\ell=1}^{L}\mu_{\ell}\left[\log_{2}1+
    \rho_{\ell}\hspace{1pt}||\mathbf{g}_{\ell}||^{2}\right], 
    \vspace{-4pt}
\end{equation}
where $\mathbf{g}_{\ell}$ is the projection of $\mathbf{h}_{\ell}$ onto the matrix null space of $\{\mathbf{h}_{1},\mathbf{h}_{2},\dots{},\mathbf{h}_{\ell-1},\mathbf{h}_{\ell+1},\dots{},\mathbf{h}_{L}\}$. Rather interestingly, the optimization problem quoted in \eqref{OptimalPowerAnalysis6} is \emph{identical} as that quoted in \eqref{OptimalPowerAnalysis4}, with the exception that $\mathbf{f}_{\ell}$ is replaced by $\mathbf{g}_{\ell}$ which does not contribute to the asymptotic solution. \emph{This only affects the $\mathcal{O}(1)$ term in \eqref{OptimalPowerAnalysis5} and thus the power allocation policy of 
$\rho_{\ell}=\mu_{\ell}\rho$ is also the asymptotic solution to \eqref{OptimalPowerAnalysis6}.} 

Using the asymptotically optimal power allocation, the weighted sum capacity of DPC can be expressed as 
\vspace{-5pt}
\begin{equation}
    \label{OptimalPowerAnalysis7}
    C_{\textrm{DPC}}\left(\mu,\mathbf{H},\rho\right)\approx\sum\limits_{\ell=1}^{L}
    \mu_{\ell}\left[\hspace{1pt}\log_{2}1+\rho\hspace{1pt}\mu_{\ell}\hspace{1pt}||\mathbf{f}_{\ell}||^{2}\hspace{1pt}\right], 
    \vspace{-5pt}
\end{equation}
while the weighted sum capacity for ZF can be written as 
\vspace{-5pt}
\begin{equation}
\vspace{-4pt}
    \label{OptimalPowerAnalysis8}
    C_{\textrm{ZF}}\left(\mu,\mathbf{H},\rho\right)\approx\sum\limits_{\ell=1}^{L}
    \mu_{\ell}\left[\log_{2}1+\rho\hspace{1pt}\mu_{\ell}\hspace{1pt}||\mathbf{g}_{\ell}||^{2}\right].  
    \vspace{4pt}
\end{equation}
Note that \eqref{OptimalPowerAnalysis7} and \eqref{OptimalPowerAnalysis8} utilize the classical affine approximation. Thus, the instantaneous sum capacity loss per-channel realization is given by 
\vspace{-6pt}
\begin{equation}
    \label{OptimalPowerAnalysis9}
    C_{\textrm{loss}}^{\textrm{DPC-ZF*}}\left(\mu,\mathbf{H},\rho\right)\approx
    \sum\limits_{\ell=1}^{L}\mu_{\ell}\left[\log_{2}\frac{||\mathbf{f}_{\ell}||^{2}}
    {||\mathbf{g}_{\ell}||^{2}}\right]. 
    \vspace{-2pt}
\end{equation}
As such, the expected capacity loss can be computed by evaluating $\mathbb{E}\{ C_{\textrm{loss}}^{\textrm{DPC-ZF*}}(\mu,\mathbf{H},\rho)\}$. In the following section, for a given channel realization, we demonstrate the negligible difference between the true weighted sum capacity and the weighted sum capacity achievable using the derived power allocation solution in Sec.~\ref{WeightedCapacityMaximization}, i.e.,  $\rho_{\ell}=\rho\hspace{1pt}\mu_{\ell}$.

\vspace{-4pt}
\section{Numerical Results}
\label{NumericalResults}
\vspace{-1pt}
We consider the setting where the BS is located at the origin of a circular cell having a radius 100 m. The users are uniformly distributed within the cell with respect to its area. We assume operation at 3.7 GHz, where ${\lambda}=$8.1 cm and employ $d=0.5\lambda$ inter-element antenna spacing at the BS array. We assume uniformly distributed LOS steering angles to each user from the BS, denoted earlier as $\phi_{m}$, within the interval $[0,2\pi]$ across the azimuthal plane. \emph{Unless otherwise stated, we assume the following parameters:} The BS is equipped with a ULA of $M=$ 64 elements, serving $L=$ 8 users for single-antenna ($N=$ 1), single-stream transmission. For multi-stream transmission, we consider service to $L=$ 4 users, each having $N=$ 2 receive antennas, such that the \emph{total number} of receive antennas across \emph{all terminals} remains consistent with the single-stream case for comparison purposes. For single-stream transmission, ZF precoding and DPC performance is evaluated, while for multi-stream, BD precoding and DPC performance is evaluated.
\begin{figure}[!t]
\vspace{-8pt}
    \includegraphics[width=8.3cm]{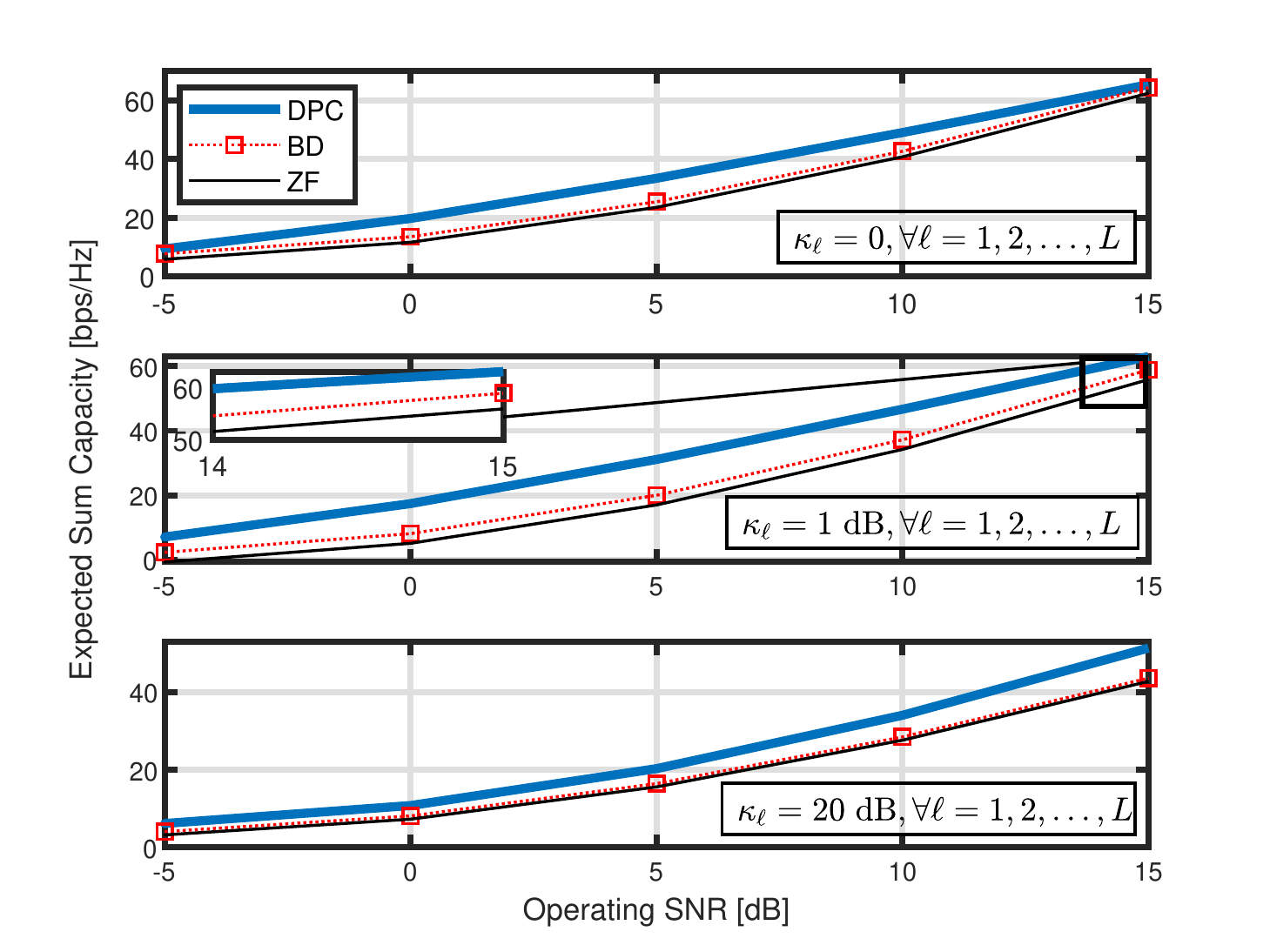}   
    \vspace{-10pt}
    \caption{Expected sum capacity vs. operating SNRs with DPC, BD and ZF processing.}
    \label{fig:ExpectedSumCapacityRayleighRiceLowHighK}
    \vspace{-20pt}
\end{figure}

Figure~\ref{fig:ExpectedSumCapacityRayleighRiceLowHighK} depicts the expected sum capacity performance as a function of operating SNRs. The top sub-figure shows the DPC  performance relative to BD and ZF, for thr baseline case of \emph{Rayleigh fading channels}. As such, $\kappa_{\ell}=0$, $\forall{}\ell=1,2,\dots{}L$. Several important trends can be noted: (1) We can observe that both ZF and BD suffer from an expected capacity loss relative to DPC across all SNRs. This is due to a penalty in the desired signal power incurred in orthogonalizing interfering channels. (2) At both high and low SNRs, the achieved capacities converge for both optimal (DPC) and sub-optimal (BD and ZF) precoding, where their performance is comparable to DPC. This is since at high SNRs, the desired power penalty is less pronounced (in magnitude) due to the scaling of the ZF/BD beamforming vectors/matrices by high operating SNRs. In contrast, at low SNRs, noise power dominates performance, and thus all techniques yield almost equal performance. (3) BD performance is marginally better than ZF due to multi-stream transmission, as also observed by \cite{LEE1}. In contrast to the top sub-figure, the middle and bottom sub-figures depict the expected sum capacity performance with \emph{fixed} $\kappa_{\ell}=$1 dB and $\kappa_{\ell}=$20 dB $\forall{}\ell=1,2,\dots{},L$. These serve as test cases denoting low and high LOS powers. The performance obtained with low LOS powers is similar to the baseline case of Rayleigh fading. However, the expected sum capacity performance is significantly \emph{lower} at higher Ricean $K$-factors across all SNRs. This is due to the \emph{ill-conditioning} of the matrix inverse for both ZF and BD, respectively. As shown in \cite{RUSEK1}, with strong LOS presence, the \emph{condition number} of the inverted matrix (ratio of the largest to smallest singular values of the composite channel covariance matrix) is \emph{larger (on the order of 20 dB on average via simulations)} relative to the case with strong NLOS conditions. This implies that strong LOS increases the spread of singular values relative to strong NLOS, inducing higher correlation levels and thereby reducing performance. Moreover, with high $K$-factors, the relative gap between DPC and ZF, BD performance at higher SNRs tends to increase, in comparison to the Rayleigh fading. 
\begin{figure}[!t]
\vspace{-10pt}
    \includegraphics[width=8.3cm]{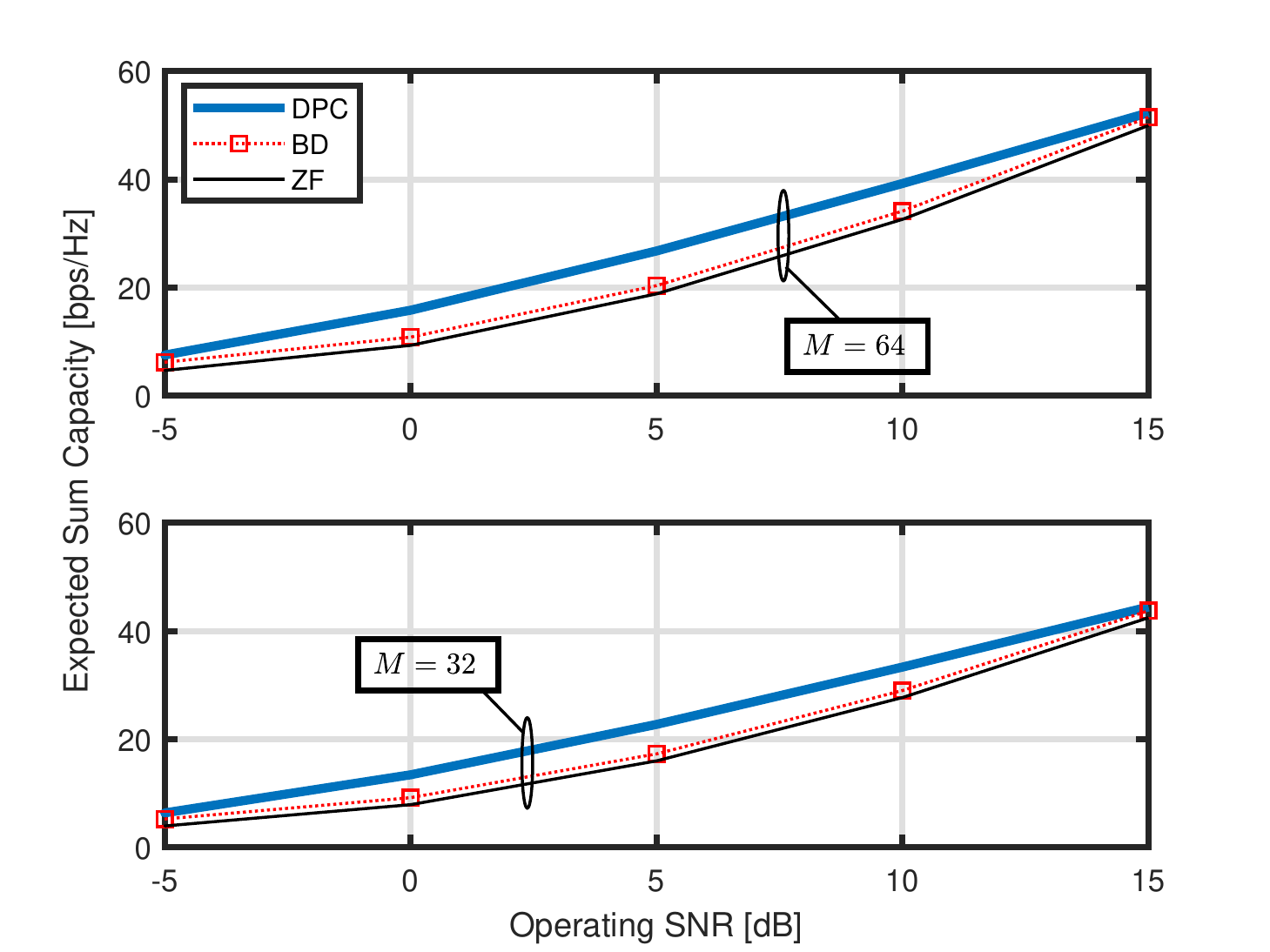}   
    \vspace{-7pt}
    \caption{Expected sum capacity vs. operating SNRs for varying $M$.}
    \label{fig:ExpectedSumCapacityRandomKVaryingM}
    \vspace{-19pt}
\end{figure}

In the general case, when each user has a specific $K$-factor, we consider the model adopted in \cite{TATARIA2}. Here,  $\kappa_{\ell}\sim{}\textrm{ln}(9,5),\forall{}\ell=1,2,\dots,L$, i.e., the $K$-factors are drawn from a lognormal distribution with mean of 9 dB and variance of 5 dB. Figure~\ref{fig:ExpectedSumCapacityRandomKVaryingM} depicts the expected sum capacity performance as a function of operating SNRs for the case where each user has a specific $K$-factor. The top sub-figure shows DPC, BD and ZF performance with $M=64$, while the bottom sub-figure shows the equivalent performance with $M=32$, where a proportional reduction in the expected sum capacities is observed, as expected. For the $M=64$ case, one can observe that the expected sum capacities yield values in between those for low and high $K$-factor cases in Figure~\ref{fig:ExpectedSumCapacityRayleighRiceLowHighK}. This is since on average, $K$-factor values close to 9 dB are drawn (between 1 dB and 20 dB), yet the lognormal nature of $K$-factor density occasionally yields larger $K$ values, skewing the result. 

\begin{figure}[!t]
\vspace{-10pt}
\hspace{3pt}
    \includegraphics[width=8.2cm]{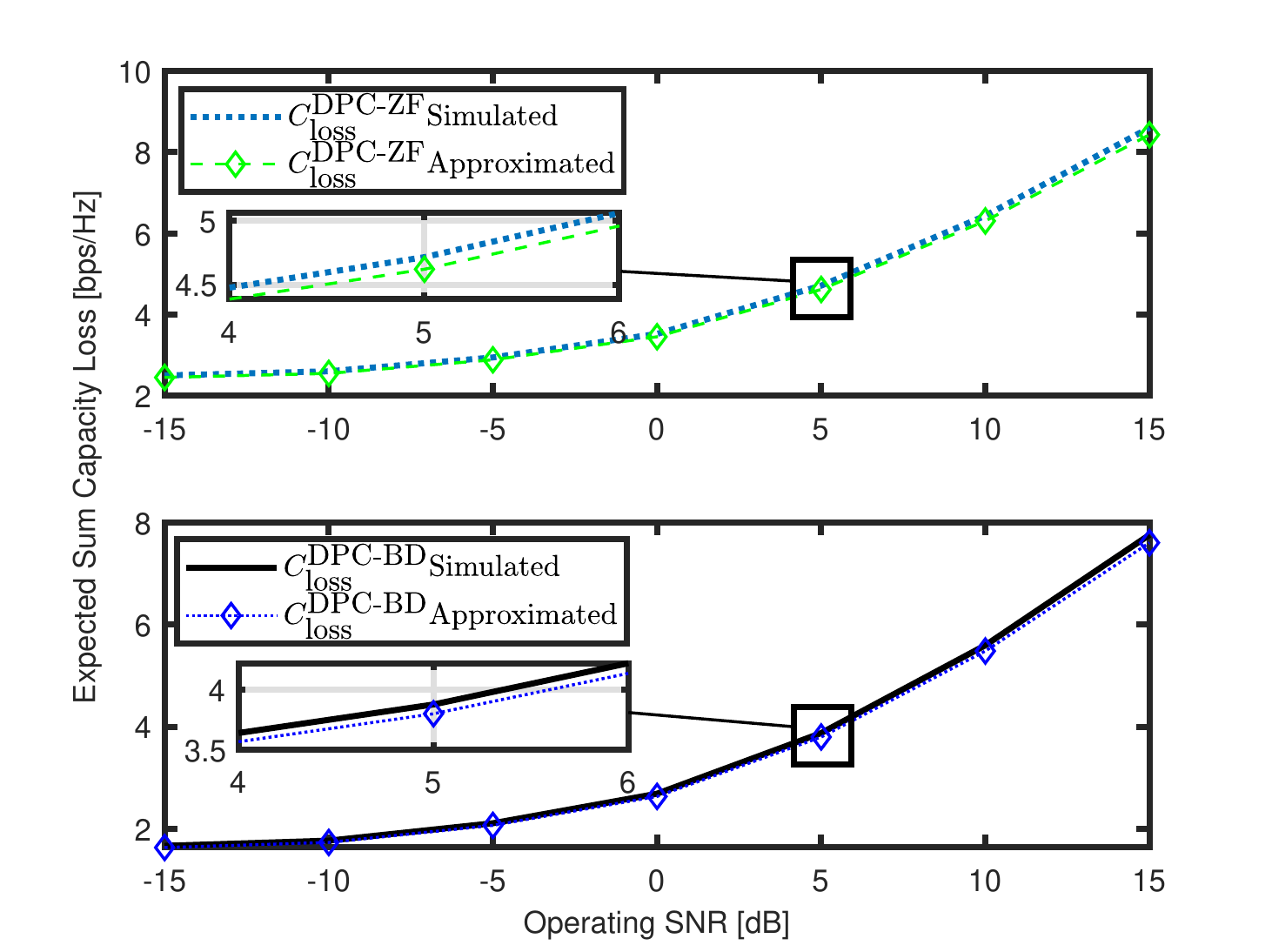}
    \vspace{-10pt}
    \caption{Expected sum capacity loss between DPC and ZF, BD precoding vs. operating SNRs.}
    \label{fig:ApproximationAccuracy}
    \vspace{-15pt}
\end{figure}
Figure~\ref{fig:ApproximationAccuracy} demonstrates the expected sum capacity loss  between DPC and ZF, as well as BD precoding when each user has $\kappa_{\ell}\sim{}\textrm{ln}(9,5),\forall{}\ell=1,2,\dots,L$. One can observe that the derived approximations of the expected capacity loss in \eqref{capacitylossanalysis1}-\eqref{capacitylossanalysis7} and  \eqref{BDCapacityAnalysis1}-\eqref{BDCapacityAnalysis3} for both ZF and BD precoding tightly match the simulated expected sum capacity loss. \emph{This confirms the generality of our analysis, as promised earlier in the paper.} 

\begin{figure}[!t]
\hspace{5pt}
    \includegraphics[width=8cm]{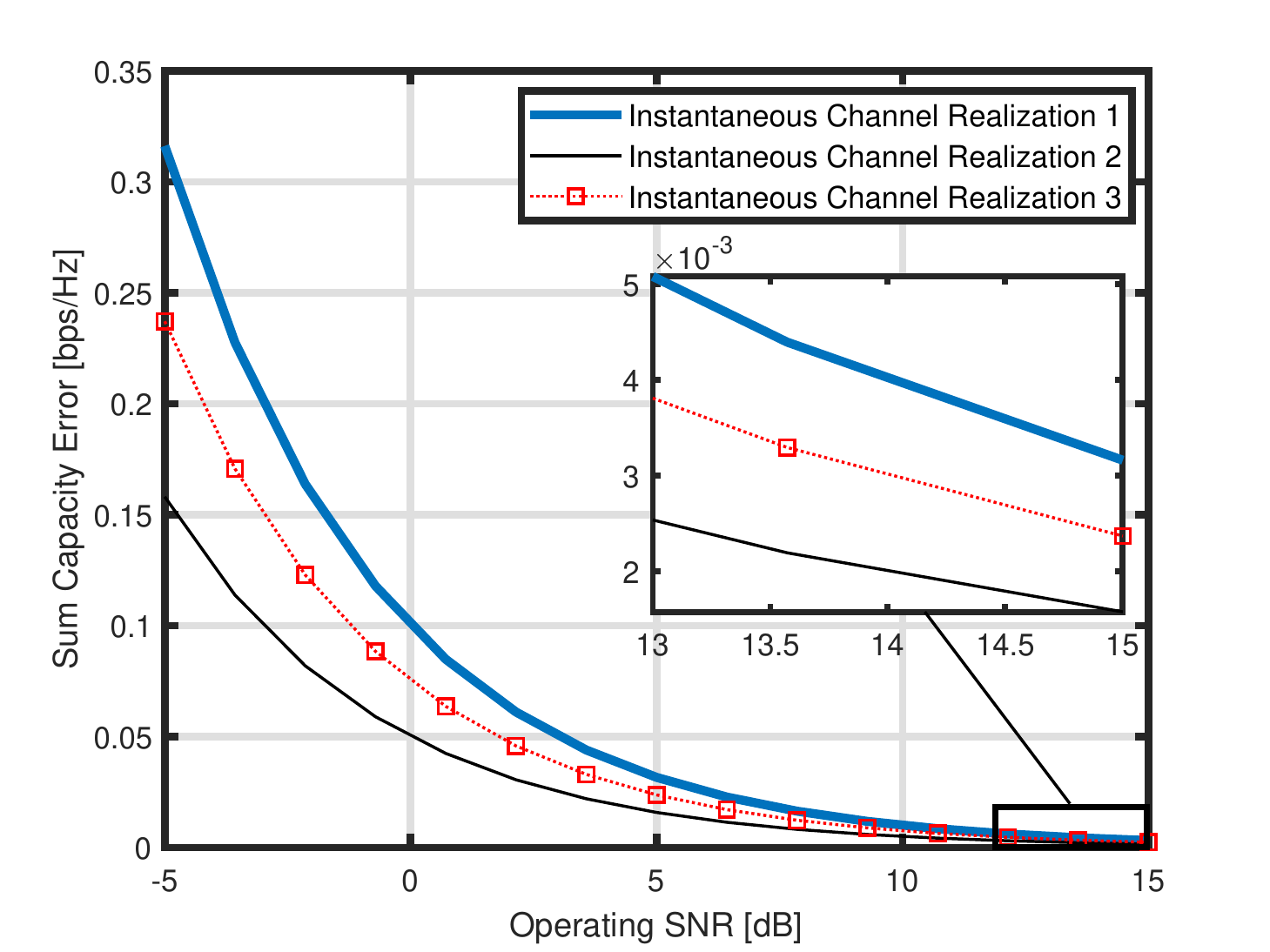} 
    \vspace{-7pt}
    \caption{Weighted sum capacity difference between the derived exact and asymptotic solutions with $N=1$, i.e., for DPC and ZF precoding.}
    \label{fig:WeightedSumCapacityDifference}
    \vspace{-18pt}
\end{figure}

Figure~\ref{fig:WeightedSumCapacityDifference} demonstrates the weighted sum capacity difference, referred to as the ``sum capacity error" on the $y$-axis of the figure, between the solution in \eqref{OptimalPowerAnalysis1} and the asymptotic solution of $\rho_{\ell}=\rho\hspace{1pt}\mu_{\ell}\hspace{1pt},\forall{}\ell=1,2,\dots,L$. Since single-stream transmission is required, the result is shown for $L=2$, $N=1$ and $M=32$ with fixed $\kappa_{1}$ and $\kappa_{2}\sim{}\textrm{ln}(9,5)$. We fix $\mu_{1}=$0.6 and $\mu_{2}=$0.4 for \emph{three instantaneous channel realizations} to both users. We can observe that the sum capacity error is negligible across the entire SNR range considered. This is interesting, since even though the derived expressions rely on high SNRs, the analysis methodology seems to be applicable over a much wider range of SNRs, demonstrating its generality.

\vspace{-3pt}
\section{Conclusions}
\label{Conclusions}
\vspace{-2pt}
We approximate the difference between the expected sum capacity achieved by optimal and sub-optimal precoding techniques of DPC, ZF and BD. Our analysis methodology utilizes the affine approximation and computes the asymptotic capacity loss for multiuser broadcast channels. Unlike previously, our analysis caters for the more general propagation conditions captured by Ricean fading and considers maximum channel heterogeneity across multiple users by considering user specific parameters. It is shown that linear precoding techniques incur a moderately high expected sum capacity loss penalty relative to DPC, yet this penalty is much smaller at high SNRs and in pure NLOS conditions. We generalized our analysis to problem of weighted sum capacity maximization. We show that allocating power proportional to user weights is asymptotically optimal at high SNR.


\end{document}